\documentclass{jfm}

\usepackage{graphicx}
\usepackage{natbib}
\usepackage{tikz,amsmath,amssymb,esint}

\newcommand{\bu}{\mathbf u}

\newcommand{\zhat}{\mathbf{\hat z}}

\newcommand{\lan}{\left\langle}
\newcommand{\ran}{\right\rangle}
\newcommand{\wT}{\lan wT\ran}

\newcommand{\Q}{\mathcal Q}
\newcommand{\F}{\mathcal F}
\newcommand{\wt}{\widetilde}
\newcommand{\Top}{h^{T}}
\newcommand{\Bot}{h^{B}}

\newcommand{\hsq}{\Vert \bnabla h\Vert^2}

\newcommand\solidrule[1][21pt]{\rule[0.5ex]{#1}{.4pt}}
\newcommand\dashedrule{\mbox{%
	\solidrule[5pt]\hspace{3pt}\solidrule[5pt]\hspace{3pt}\solidrule[5pt]}}

\title{Bounds for convection between\\rough boundaries}

\author{David Goluskin\aff{1}\corresp{\email{goluskin@umich.edu}} and Charles R.\ Doering\aff{1,2}}

\affiliation{\aff{1} Department of Mathematics and Center for the Study of Complex Systems, University of Michigan, Ann Arbor, MI 48109, USA
	\aff{2} Department of Physics, University of Michigan, Ann Arbor, MI 48109, USA}

\shorttitle{Bounds for convection between rough boundaries}
\shortauthor{D.\ Goluskin \& C.\ R.\ Doering}

\begin{document}

\maketitle

\begin{abstract}
We consider Rayleigh--B\'enard convection in a layer of fluid between rough no-slip boundaries where the top and bottom boundary heights are functions of the horizontal coordinates with square-integrable gradients. We use the background method to derive an upper bound on mean heat flux across the layer for all admissible boundary geometries. This flux, normalized by the temperature difference between the boundaries, can grow with the Rayleigh number ($Ra$) no faster than ${\cal O}(Ra^{1/2})$ as $Ra \rightarrow \infty$. Our analysis yields a family of similar bounds, depending on how various estimates are tuned, but every version depends explicitly on the boundary geometry. In one version the coefficient of the ${\cal O}(Ra^{1/2})$ leading term is $0.242 + 2.925\hsq$, where $\hsq$ is the mean squared magnitude of the boundary height gradients. Application to a particular geometry is illustrated for sinusoidal boundaries.
\end{abstract}


\section{Introduction}

Averaged transport properties of turbulent fluid flows, such as mixing efficiencies or fluxes of momentum and heat, are often of more interest than details of any particular realization. One way to estimate mean quantities is to numerically integrate the equations of motion over a long time and average. Of course the broad range of scales present in turbulence makes such direct numerical simulations challenging. An alternate approach is to derive mathematical bounds on mean quantities. Although such bounds might not be as close to ``true" values as the results of simulations, they have certain advantages. First, bounds are proven directly on infinite-time averages whereas simulations may be subject to long transients. Second, bounds apply to all possible initial conditions whereas an attracting state reached in one simulation does not preclude the existence of other attracting states. Finally, bounds may be derived as functions of the relevant parameters and, as in the present work, for a variety of geometries within some class. In contrast, simulations must be repeated for every new parameter value or geometry, and they often cannot access the extreme parameter regimes relevant for many astrophysical, geophysical, and engineering applications.

Mathematical methods for bounding turbulent transport properties have largely focused on the simple geometry of a fluid layer confined by smooth planar walls. For channels and shear flows, upper bounds on wall drag coefficients have been derived rigorously from the incompressible Navier-Stokes equations \citep{Busse1969a, Howard1972, Constantin1994, Doering1995a, Kerswell1997, Nicodemus1998, Hoffmann1999, Kerswell2002, Plasting2003, Seis2015}. For buoyancy-driven thermal convection between planar boundaries, the authors and others have proven lower bounds on mean temperature \citep{Lu2004, Whitehead2011, Whitehead2012, Goluskin2015a, Goluskin2015} and upper bounds on heat transport \citep{Howard1963, Constantin1996, Kerswell1997, Kerswell2001, Otero2002, Plasting2005, Wittenberg2010, Otto2011a, Whitehead2011a, Whitehead2012, Wen2013, Wang2013, Whitehead2014, Choffrut2016}.

Domain geometries in applications are often more complicated than a plane layer, however, and physical boundaries are never perfectly smooth. Such boundary variations can significantly affect mean quantities of interest, so closing the gap between theory and experiment requires establishing bounds that are valid for non-uniform, geometrically complicated or rough boundaries. Here we present a way to do so using the background method, a variational method used to prove many of the above-cited bounds in simple geometries. Complicated geometries have been studied previously by \citet{Wang1997}, who bounded dissipation in certain shear-driven flows, albeit without computing explicit constants. Here we study thermal convection by a somewhat different approach that produces explicit bounds.

The particular problem we consider in this paper is Rayleigh--B\'enard convection where a layer of fluid heated from below and cooled from above is confined between non-intersecting top and bottom boundaries that are no-slip and perfectly conductive (that is, isothermal). Unlike the canonical configuration of \citet{Rayleigh1916} in which the cooler upper and warmer lower boundaries are perfectly flat planes, we allow the boundary heights to be continuous, piecewise differentiable functions of the horizontal coordinates. As previously accomplished for planar boundaries, we derive an upper bound on the net vertical heat flux across the fluid layer as a function of the temperature difference between the boundaries and the relevant material and geometric parameters. Such bounds are typically expressed in terms of the dimensionless Nusselt and Rayleigh numbers, $Nu$ and $Ra$, where $Nu$ is the factor by which convection amplifies heat transfer beyond diffusive transport, and $Ra$ is proportional to the temperature difference across the layer. For finite-Prandtl-number fluids between no-slip \emph{planar} boundaries, the best upper bound proven to date is $Nu\lesssim0.027\,Ra^{1/2}$ \citep{Plasting2003}.

In this work we prove an upper bound on heat transport that also scales proportionally to $Ra^{1/2}$, albeit with a prefactor that depends on the details of the boundary geometry. This result is consistent with past experimental and numerical studies of convection between rough boundaries \citep{Du1998, Du2000, Villermaux1998, Ciliberto1999, Roche2001, Stringano2006, Wei2014, Salort2014, Wagner2015, Toppaladoddi2015}, all of which report that roughness can enhance heat transport, and some of which report that it increases the rate at which $Nu$ grows with $Ra$. The fastest growth rate reported is $Ra^{1/2}$ in the experiments of \citet{Roche2001} where all boundaries were rough, including the side walls.

Section \ref{sec: config} defines the mathematical model we study, and \S\ref{sec: int quant} lays out key integral relations and the meanings of various integral quantities, including the Nusselt number. The main result is proven in \S\ref{sec: proof}, and its application to a particular geometry is illustrated for sinusoidal boundaries. Section \ref{sec: con} offers conclusions and open challenges.


\section{Configuration \label{sec: config}}

We employ the Oberbeck-Boussinesq approximation with constant kinematic viscosity $\nu$, thermal diffusivity $\kappa$, and coefficient of thermal expansion $\alpha$. We nondimensionalize such that the length scale $d$ is the difference between the maximum height of the top boundary and the minimum height of the bottom one, the time scale $d^2/\kappa$ is that of thermal diffusion, and the temperature scale $\Delta$ is the temperature difference between the top and bottom boundaries. Then the dimensionless Boussinesq equations governing the velocity $\mathbf{u}=(u,v,w)$, temperature $T$, and pressure $p$ are 
\begin{align}
\mathbf u_t + \mathbf u \bcdot \bnabla \mathbf u  &= 
-\bnabla p + \Pran \bnabla^2 \mathbf u + \Pran Ra\,T \mathbf{\hat z}, \label{eq: u} \\
T_t + \mathbf u \bcdot \bnabla T& = \bnabla^2 T, \label{eq: T} \\
\bnabla \bcdot \mathbf u &= 0. \label{eq: inc}
\end{align}
The Rayleigh and Prandtl numbers are $Ra = \frac{g\alpha d^3\Delta}{\kappa\nu}$ and $\Pran = \frac{\nu}{\kappa}$, where $g$ is the acceleration of gravity in the $-\mathbf{\hat z}$ direction. Here $z$ is the vertical coordinate, while $x$ and $y$ are horizontal. Subscripts of $t$, $x$, $y$, or $z$ denote partial derivatives.

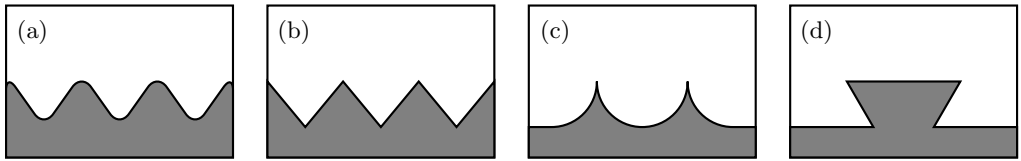
\begin{figure}
\begin{center}
\begin{tikzpicture}
\draw[thick] (0,0) rectangle (3,2);
\draw (.35,1.65) node {(a)};
\filldraw[fill=gray, thick] 
	(3,0) -- (0,0) 
	[rounded corners=6pt] -- (0,1.1) -- (.5,.4) -- (1,1.1) -- (1.5,.4) -- (2,1.1) -- 
	(2.5,.4) -- (3,1.1) [sharp corners] -- cycle;
\end{tikzpicture}
\hspace{6pt}
\begin{tikzpicture}
\draw[thick] (0,0) rectangle (3,2);
\draw (.35,1.65) node {(b)};
\filldraw[fill=gray, thick] 
	(3,0) -- (0,0) -- (0,1) -- (.5,.4) -- (1,1) -- (1.5,.4) -- (2,1) -- (2.5,.4) -- (3,1) -- cycle;
\end{tikzpicture}
\hspace{6pt}
\begin{tikzpicture}
\draw[thick] (0,0) rectangle (3,2);
\draw (.35,1.65) node {(c)};
\filldraw[fill=gray, thick] 
	(3,0) -- (0,0) -- (0,.4) -- (.3,.4) arc (270:360:.6) arc (180:270:.6) arc (270:360:.6) arc (180:270:.6) -- (3,.4) -- cycle;
\end{tikzpicture}
\hspace{6pt}
\begin{tikzpicture}
\draw[thick] (0,0) rectangle (3,2);
\draw (.35,1.65) node {(d)};
\filldraw[fill=gray, thick] 
	(3,0) -- (0,0) -- (0,.4) -- (1.1,.4) -- (.75,1) -- (2.25,1) -- (1.9,.4) -- (3,.4) -- cycle;
\end{tikzpicture}
\end{center}
\caption{\label{fig: examples} Examples of boundaries to which our main result does and does not apply. It applies to boundaries specified by continuous and piecewise differentiable functions whose gradients are (a) continuous or (b) piecewise continuous, including (c) gradients that are unbounded but square-integrable, but not to (d) boundaries that are not functions of the horizontal coordinates.}
\end{figure}

The top and bottom boundaries of the domain are specified by $z=\Top (x,y)$ and $z=\Bot (x,y)$: continuous functions that can model both small-scale roughness and large-scale features. Our proof relies on the boundaries being continuous, piecewise differentiable functions of the horizontal coordinates and on the gradients $\bnabla \Top $ and $\bnabla \Bot $ being square-integrable. Examples of boundary geometries to which our main result does and does not apply are shown, respectively, in figures \ref{fig: examples}(a-c) and \ref{fig: examples}(d).

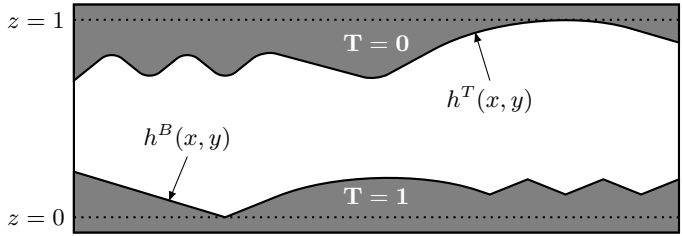
\begin{figure}
\begin{center}
\begin{tikzpicture}
\draw[thick] (0,0) rectangle (8,3);
\filldraw[fill=gray, thick] 
	(8,0) -- (0,0) -- (0,.8) -- (2,.2) 
 	[rounded corners=40pt] -- (4,1)
	[sharp corners] -- (5.5,.5) -- (6,.7) -- (6.5,.5) -- (7,.7) -- 
	(7.5,.5) -- (8,.7) -- cycle;
\filldraw[fill=gray, thick] 
	(8,3) -- (0,3) -- (0,2) 
 	[rounded corners=5pt] -- (.5,2.4) -- (1,2) -- (1.5,2.4) -- (2,2) -- (2.5,2.4) -- (4,2)
	[rounded corners=40pt] -- (6,3.08) 
	[sharp corners] -- (8,2.5) -- cycle;
\draw[black, dotted, thick] (0,.2) -- (8,.2);
\draw[black, dotted, thick] (0,2.8) -- (8,2.8);
\draw (-.5,.2) node {$z=0$};
\draw (-.5,2.8) node {$z=1$};
\draw (1.5,1.25) node {$\Bot (x,y)$};
\draw (5.5,1.75) node {$\Top (x,y)$};
\draw[-latex] (1.5,1) -- (1.25,.4);
\draw[-latex] (5.5,2) -- (5.3,2.65);
\draw[white] (4,.48) node {$\mathbf{T=1}$};
\draw[white] (4,2.5) node {$\mathbf{T=0}$};
\end{tikzpicture}
\end{center}
\caption{\label{fig: setup} Example two-dimensional slice of our nondimensionalized configuration}
\end{figure}

Lengths are nondimensionalized so that the maximum of $\Top (x,y)$ and the minimum of $\Bot (x,y)$ differ by unity. We choose $z=0$ to coincide with the minimum of $\Bot (x,y)$, hence the vertical domain is
\begin{equation}
0\le \Bot (x,y)\le z\le \Top (x,y)\le1.
\end{equation}
Figure \ref{fig: setup} is a schematic of the setup we have in mind.
The origin of the temperature scale is arbitrarily, so for the thermal boundary conditions we can fix
\begin{align}
T\big|_{z=\Bot (x,y)}&=1, &
T\big|_{z=\Top (x,y)}&=0. 
\label{eq: T BC}
\end{align}

When the boundaries are planar, standard Rayleigh--B\'enard convection is recovered with $\Bot \equiv0$ and $\Top \equiv1$. The horizontal coordinates $(x,y)$ are assumed to lie in a bounded two-dimensional domain $\Omega$ that need not be simple. In each horizontal direction the flow can be either periodic (with $\Bot $ and $\Top $ also periodic) or bounded by perfectly insulating vertical sides. If there are no periodic directions, our three-dimensional domain is a cylinder with arbitrary cross-section $\Omega$ and rough end caps defined by $\Bot$ and $\Top$. Consideration of more complicated side walls is left for future work. We impose no-slip ($\bu=\mathbf0$) conditions on the top and bottom boundaries. If side walls exist, velocity conditions there can be either no-slip or free-slip. The existence of piecewise smooth solutions to the three-dimensional Boussinesq equations is unproven even in simple geometries, but our results apply to such solutions whenever they exist.

Unless there is no buoyancy ($Ra=0$), the configuration we are considering does not generally admit a static state -- that is, a solution to the equations of motion with $\bu=\mathbf0$ and $T_t=0$. This is distinct from the case of planar boundaries where the static solution exists for all $Ra$ and is stable when $Ra<1707.7$ \citep{Jeffreys1928}. Indeed, a static state must satisfy $\bnabla^2T=0$ and $\bnabla p=\Pran Ra\,T \, \mathbf{\hat z}$. Such a stationary harmonic temperature field generally exists but the buoyancy force field it induces can only be balanced by the pressure if $Ra\,T \, \mathbf{\hat z}$ is a gradient -- that is, only if $Ra=0$ or if $\Bot$ and $\Top$ are constant so that the resulting $T$ is a function of $z$ alone.


\section{Integral quantities \label{sec: int quant}}

We wish to bound the dimensionless time-averaged heat flux across the layer, $\F$.
For concreteness we define $\F$ as the inward heat flux across the bottom boundary, although it follows from the temperature equation (\ref{eq: T}) that the time-averaged heat flux is the same across any surface within the fluid that spans the domain horizontally. At each point on the bottom boundary, the inward heat flux is equal to the outward temperature gradient, $\mathbf{\hat n}\bcdot\bnabla T$. Integrating over the bottom surface and averaging over infinite time gives the definition
\begin{equation}
\F := \lim_{t\to\infty}\frac{1}{t}\int_0^tdt'\int\limits_{\{z=\Bot \}} dS\, \mathbf{\hat n}\bcdot\bnabla T(x,y,z,t').
	\label{eq: F}
\end{equation}
Here we assume that infinite-time averages exist, but this assumption could be avoided by employing $\limsup$ in place of $\lim$. The main result we establish in \S\ref{sec: proof} is an upper bound on $\F$ that scales like $Ra^{1/2}$ at large $Ra$.

Let angular brackets denote volume integrals and infinite-time averages:
\begin{equation}
\lan f\ran := \lim_{t\to\infty}\frac{1}{t}\int_0^tdt'\iint_\Omega dxdy 
	\int_{\Bot (x,y)}^{\Top (x,y)}dz\, f(x,y,z,t').
\label{eq: angular}
\end{equation}
Our calculations are expressed in terms of volume integrals, as opposed to volume averages, so enlarging the horizontal extent tends to increase $\F$. Integrating $\lan T\times(\ref{eq: T})\ran$ and $\lan\bu\bcdot(\ref{eq: u})\ran$ by parts yields the integral relations
\begin{align}
\lan|\bnabla T|^2\ran &= \F, \label{eq: T power int} \\
\lan|\bnabla\bu|^2\ran &= Ra\wT. \label{eq: u power int}
\end{align}
Time derivatives do not appear in these relations because the volume integrals of $|\bu|$ and $|T|$ are bounded uniformly in time, a fact that follows from the present analysis \citep[cf.][]{Constantin1996}.

The heat flux $\F$ that we want to bound is related \emph{a priori} to the thermal dissipation $\lan|\bnabla T|^2\ran$ but \emph{not} to the viscous dissipation $\lan|\bnabla\bu|^2\ran$. With planar boundaries $\F$ is related also to $\lan|\bnabla\bu|^2\ran$ because it is tied to the mean vertical convection $\wT$ by $\F=A+\wT$, where $A=|\Omega|$ is the area of the horizontal cross-section. With non-planar boundaries, on the other hand, a given flux $\F$ can coincide with various $\wT$, depending on how the heat transport is distributed horizontally; if a larger fraction of convective transport occurs in regions where the layer is thin, $\wT$ will be smaller, essentially because conduction within the boundaries helps more with upward heat transport.

Implementations of the background method for Rayleigh--B\'enard convection with planar boundaries have exploited the fact that $\F$ is related to both $\lan|\bnabla T|^2\ran$ and $\lan|\bnabla \bu|^2\ran$ \citep[e.g.][]{Constantin1996, Kerswell2001}. Since non-planar boundaries spoil this feature, our implementation is more similar to lower bounds on mean temperature proven for internally heated convection. Mean temperature in such configurations, like $\F$ here, is related \emph{a priori} to $\lan|\bnabla T|^2\ran$ but not to $\lan|\bnabla \bu|^2\ran$ \citep{Lu2004, Whitehead2011, Whitehead2012, Goluskin2015a, Goluskin2015}.

Heat transport in convection is typically quantified by a Nusselt number $Nu$ that is in some sense a ratio of total transport to conductive transport. Defining both transports as volume averages in the flowing fluid has proven useful for revealing parallels between various thermal boundary conditions \citep{Otero2002, Johnston2009, Wittenberg2010, Goluskin2015}. However, it is unclear how to extend this definition to complicated geometry, partly because the convective transport $\wT$ is no longer related \emph{a priori} to total transport. The natural alternative is to normalize $\F$ by the heat flux in the static state that exists when $Ra=0$. Defined in this way, $Nu$ is unity in the static state and is suitably insensitive to the domain size. Furthermore, the bound we prove on $\F$ implies an upper bound on $Nu$ that scales proportionally to $Ra^{1/2}$, as in the case of planar boundaries.
 

\section{Upper bound on heat transport \label{sec: proof}}

Rather than bound $\F$ subject to the constraint that $\bu$ and $T$ solve the Boussinesq equations (\ref{eq: inc})--(\ref{eq: T}), we bound $\F$ subject to relaxed constraints on $\bu$ and $T$ that are satisfied by all such solutions. In particular, the only constraints we impose are incompressibility, boundary conditions on $\bu$ and $T$, and three integral relations implied by the Boussinesq equations. The analysis of this section culminates in an explicit bound on $\F$ for general boundaries in \S\ref{sec: main result}, followed by its application to sinusoidal boundaries in \S\ref{sec: sinusoidal}.


\subsection{\label{sec: decomp}Background decomposition}

The background method is implemented by decomposing the temperature field as
\begin{equation}
T(x,y,z,t) = \tau(x,y,z) + \theta(x,y,z,t),
\label{eq: decomp}
\end{equation}
where $\tau$ is the so-called background field. We choose an explicit expression for $\tau$ that is continuous and satisfies the same boundary conditions as $T$, so $\theta$ is continuous and satisfies homogenous boundary conditions.

The first two integral constraints we require are the dissipation balances \eqref{eq: T power int} and \eqref{eq: u power int}. The last, obtained by integrating $\lan\theta\times\eqref{eq: T}\ran$ by parts, is
\begin{equation}
\lan\bnabla\tau\bcdot\bnabla\theta\ran = -\lan\theta\bu\bcdot\bnabla\tau\ran - \lan|\bnabla\theta|^2\ran.
\label{eq: relation 3}
\end{equation}
Our only use for these integral constraints is to derive a particular equality for $\F$:
\begin{align}
\F =& \lan|\bnabla \tau|^2\ran + 2\lan\bnabla\tau\bcdot\bnabla\theta\ran +
	\lan|\bnabla \theta|^2\ran \label{eq: F 1} \\
	=& \lan|\bnabla \tau|^2\ran -2 \lan\theta\bu\bcdot\bnabla\tau\ran -
	\lan|\bnabla \theta|^2\ran \label{eq: F 2} \\
	=& \lan|\bnabla \tau|^2\ran +
	 \left[a\lan\tau w\ran - \tfrac{a}{Ra}\lan|\bnabla \bu|^2\ran - 
	 \lan\theta\bu\bcdot(2\bnabla\tau-a\zhat)\ran - \lan|\bnabla \theta|^2\ran \right],
	\label{eq: F 3}
\end{align}
where $a>0$ remains to be specified. The first line above is the expansion of (\ref{eq: T power int}) in terms of $\tau$ and $\theta$, the next follows from (\ref{eq: relation 3}), and the third is reached by adding $a\left[\lan w(\tau+\theta)\ran-\tfrac{1}{Ra}\lan|\bnabla\bu|^2\ran\right]$, which is zero according to (\ref{eq: u power int}).

If the boundaries were planar we could choose $\tau$ to make the part of \eqref{eq: F 3} in brackets nonpositive, thereby proving $\F \le \lan|\bnabla \tau|^2\ran$. However, doing so relies partly on choosing a horizontally uniform $\tau$ to make $\lan\tau w\ran$ vanish, and with rough boundaries this would violate the boundary conditions on $\tau$. Instead we can prove an upper bound on the bracketed expression that is larger than zero but of the same order in $Ra$ as the $\lan|\bnabla \tau|^2\ran$ term. We will do this in two pieces, splitting the $\lan|\bnabla \bu|^2\ran$ term to obtain
\begin{equation}
\F = \lan|\bnabla \tau|^2\ran + \Q_1 + \Q_2,
\label{eq: F final}
\end{equation}
where
\begin{align}
\Q_1[\bu,\theta] &:= -\left[(1-\gamma)\tfrac{a}{Ra}\lan|\bnabla \bu|^2\ran + 
	 \lan\theta\bu\bcdot(2\bnabla\tau-a\zhat)\ran + \lan|\bnabla \theta|^2\ran\right],
\label{eq: Q1} \\
\Q_2[\bu] &:= a\lan\tau w\ran - \gamma\tfrac{a}{Ra}\lan|\bnabla\bu|\ran^2,
\label{eq: Q2}
\end{align}
and $\gamma \in (0,1)$ will be chosen later.

We will construct a background field $\tau$ for which each term in \eqref{eq: F final} is no larger than ${\cal O}(Ra^{1/2})$. Regarding $\Q_1$ and $\Q_2$ as quadratic functionals of unknown fields $\theta$ and $\bu$, we prove upper bounds that hold for all $\theta$ and incompressible $\bu$ that satisfy the boundary conditions.


\subsection{\label{sec: background} Choice of background field}

The continuous background field $\tau(x,y,z)$ will be defined piecewise in three regions: top and bottom boundary layers of uniform height $\delta$, and the remaining interior. The boundary layers must be thin enough to not overlap; planar boundaries require $\delta<1/2$, and non-planar boundaries require yet smaller $\delta$. The $\tau$ considered here has affine $z$-dependence in all three regions, which leads to tractable analysis. A more general ansatz would yield a tighter bound but not necessarily a different scaling than $Ra^{1/2}$.

\begin{figure}
\centerline{\begin{tikzpicture}
\node [anchor=south west] at (0,.18) 
	{\includegraphics[height=115pt]{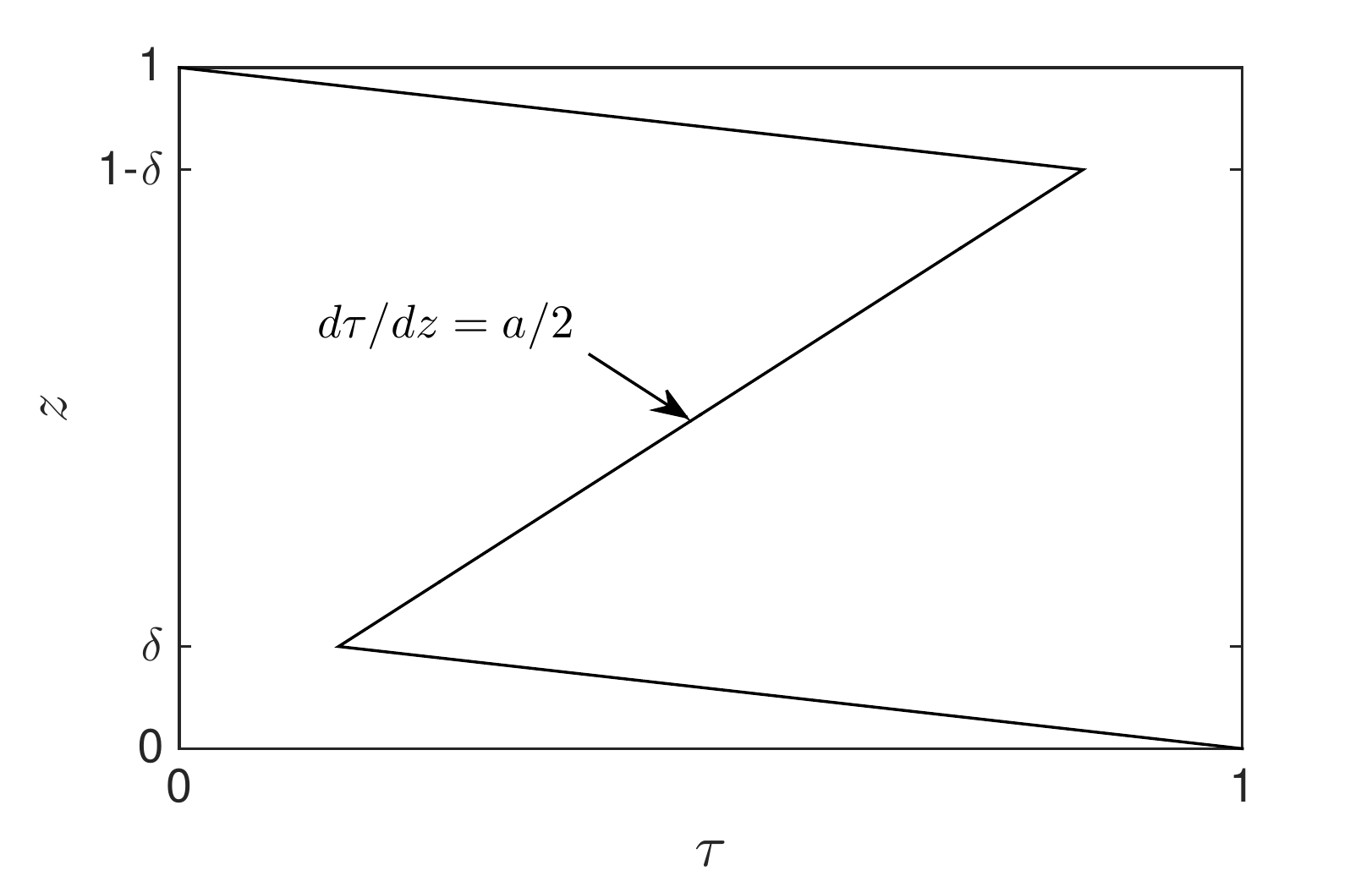}};
\node [anchor=south west] at (6.3,0) 
	{\includegraphics[height=130pt]{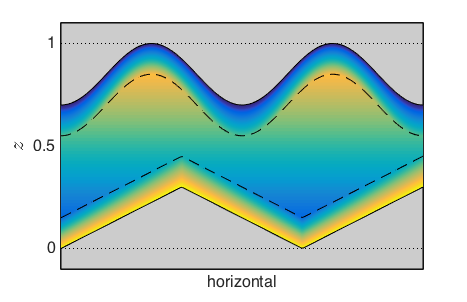}};
\draw (.2,4.3) node {(a)};
\draw (6.6,4.3) node {(b)};
\end{tikzpicture}}
\caption{\label{fig: tau} Examples of the chosen background field $\tau$ with $a=2$ and boundary layer thickness $\delta=0.15$. Panel (a) shows a graph of $\tau(z)$ for planar boundaries. Panel (b) shows a two-dimensional slice of $\tau(x,y,z)$ for non-planar boundaries, with $\tau$ ranging from 1 (light) on the bottom boundary to 0 (dark) on the top one. The $\tau(x,y,z)$ field is continuous everywhere, but its gradient is not generally continuous across the boundary layer edges (\protect\dashedrule).}
\end{figure}

Figure \ref{fig: tau} shows the structure of the background field $\tau$. On the interior region, $\tau=1/2$ on the $z=1/2$ plane and $\bnabla\tau=(0,0,a/2)$. This $\bnabla\tau$ makes the sign-indefinite term of $\Q_1$ vanish pointwise outside the boundary layers, which helps us show in \S\ref{sec: Q1} that $\Q_1\lesssim {\cal O}(Ra^{1/2})$. The definition of $\tau$ on the interior dictates values at the edge of each boundary layer that depend on $\Top $ and $\Bot $, and thus on $x$ and $y$. In each boundary layer we let $\tau$ have an affine dependence on $z$ that interpolates between $\tau$ at the boundary layer's edge and at the domain boundary. Altogether this means
\begin{equation}
\tau(x,y,z) = \begin{cases}
\left[ \tfrac{1}{2}+\tfrac{a}{2}(\Top -\tfrac{1}{2}-\delta) \right]
	\left(\frac{\Top -z}{\delta}\right) & \Top -\delta\le z\le \Top  \\
\tfrac{1}{2}+\tfrac{a}{2}\left(z-\tfrac{1}{2}\right) & 
	\Bot +\delta\le z\le \Top -\delta \\
1-\left[ \tfrac{1}{2}+\tfrac{a}{2}(\tfrac{1}{2}-\Bot -\delta) \right] 
	\left(\frac{z-\Bot }{\delta}\right) & 
	\hspace{17pt}\Bot \le z\le \Bot +\delta.
\label{eq: tau}
\end{cases}
\end{equation}
The background field depends only on $z$ in the interior but generally on all three coordinates in the boundary layers. However, $\tau$ is not fully defined until the parameters $a$ and $\delta$ are chosen. In \S\ref{sec: Q1} a particular ${\cal O}(Ra^{-1/2})$ expression for $\delta$ is chosen to ensure $\Q_1\lesssim {\cal O}(Ra^{1/2})$, and the choice of $a$ is discussed in \S\ref{sec: main result}.


\subsection{\label{sec: Q1}Upper bound on $\Q_1$}

In this subsection the quadratic functional $\Q_1[\bu,\theta]$ defined by (\ref{eq: Q1}) is bounded above by an ${\cal O}(Ra^{1/2})$ expression. For the upper bound to be independent of $\bu$ and $\theta$, the sign-indefinite term $\lan\theta\bu\bcdot(2\bnabla\tau-a\zhat)\ran$ must be estimated using the sign-definite terms $\lan|\bnabla \bu|^2\ran$ and $\lan|\bnabla \theta|^2\ran$. Expanding the indefinite term gives
\begin{equation}
\left|\lan\theta\bu\bcdot(2\bnabla\tau-a\zhat)\ran\right| \le \lan|2\tau_x u\theta|\ran +
	\lan|2\tau_y v\theta|\ran + \lan|(2\tau_z-a)w\theta|\ran.
\label{eq: indef}
\end{equation}
All three integrands in the righthand expression vanish outside the boundary layers, so only the integrals over the boundary layers must be estimated. Let us consider the first and third terms. (The second term is handled like the first.)

The derivatives of $\tau$ appearing in \eqref{eq: indef} are bounded pointwise in the bottom boundary layer ($\Bot\le z\le \Bot+\delta$) by
\begin{align}
\left|2\tau_z -a\right| 
&= \tfrac{1}{\delta}\left| 1+a\left(\tfrac{1}{2}-\Bot\right) \right| 
\le\tfrac{a+2}{2\delta}, \label{eq: dz ineq} \\
\left|2\tau_x \right|
&= \tfrac{1}{\delta}\left|1+ a\left(\tfrac{1}{2}-\Bot\right)	
	- a\left(\Bot+\delta-z\right)\right| \left|
	\Bot_x\right|
\le \tfrac{a+2}{2\delta}|\Bot_x |. \label{eq: dx ineq}
\end{align}
The above inequalities hold for sufficiently small $\delta$, and for the $a$ we choose below they hold for all admissible $\delta$. The estimates \eqref{eq: dz ineq}-\eqref{eq: dx ineq} and their analogues in the top boundary layer give
\begin{align}
\lan|(2\tau_z-a)w\theta|\ran 
&\le \left(\tfrac{a+2}{2\delta}\right) \left[\lan |w\theta|\ran_B+\lan |w\theta|\ran_T\right],
\label{eq: third term} \\
\lan|2\tau_x u\theta|\ran 
&\le \left(\tfrac{a+2}{2\delta}\right)
	\left[\lan|\Bot_x u\theta|\ran_B+\lan |\Top_x u\theta|\ran_T\right],
\label{eq: first term 2} 
\end{align}
where $\lan\bcdot\ran_B$ and $\lan\bcdot\ran_T$ are defined like \eqref{eq: angular} but with volume integrals only over the bottom and top boundary layers -- that is, the $z$-integrals are restricted to $\Bot\le z\le\Bot+\delta$ and $\Top-\delta\le z\le\Top$, respectively. Bounding the righthand terms of \eqref{eq: indef} using \eqref{eq: third term}, \eqref{eq: first term 2}, and the $y$-analogue of \eqref{eq: first term 2} gives
\begin{multline}
\left|\lan\theta\bu\bcdot(2\bnabla\tau-a\zhat)\ran\right| \le \left(\tfrac{a+2}{2\delta}\right)
	\Big(\Big[\lan |\Bot_x u\theta|\ran_B+\lan |\Top_x u\theta|\ran_T\Big] +
	\left[\lan |\Bot_y v\theta|\ran_B+\lan |\Top_y v\theta|\ran_T\right] \\
	+ \Big[\lan |w\theta|\ran_B+\lan |w\theta|\ran_T\Big] \Big).
\label{eq: indef 2}
\end{multline}

The thinness of the boundary layers can be exploited to show that the above integrals over boundary layers are small. We do this differently for the terms involving $w$ than for those involving $u$ or $v$. In the bottom boundary layer, the Cauchy--Schwarz inequality and the fact that $\theta$ vanishes at the boundary give
\begin{equation}
\vert \theta(x,y,z,t)\vert = \left\vert \int_{\Bot }^zdz'\,\theta_{z'}(x,y,z',t) \right\vert 
\le (z-\Bot )^{1/2}\left( \int_{\Bot }^{\Bot +\delta}dz
	\,\theta_z^2 \right)^{1/2},
\label{eq: w pointwise}
\end{equation}
and with no-slip boundaries the same estimates hold with $u$, $v$, or $w$ in place of $\theta$.

The $\lan|w\theta|\ran_B$ term can be estimated in the same way as when the boundaries are planar \citep[cf.][]{Constantin1996}, using the pointwise estimate \eqref{eq: w pointwise} and its analogue for $w$ to find
\begin{align}
\lan|w\theta|\ran_B 
&\le \tfrac{\delta^2}{2}\lan w_z^2\ran_B^{1/2}
	\lan\theta_z^2\ran_B^{1/2} 
\label{eq: w theta B first} \\
&\le \tfrac{\delta^2}{4}\left[ c\delta\lan w_z^2\ran_B +
	\tfrac{1}{c\delta}\lan\theta_z^2\ran_B \right]
\label{eq: w theta B}
\end{align}
for all $c>0$, where the second line follows from Young's inequality. Adding \eqref{eq: w theta B} to its counterpart holding in the top boundary layer and then extending integrals to the entire volume yields
\begin{equation}
\lan|w\theta|\ran_B + \lan|w\theta|\ran_T \le 
	\tfrac{\delta^2}{4}\left[ c\delta\lan w_z^2\ran +  \tfrac{1}{c\delta}\lan\theta_z^2\ran \right].
\label{eq: w theta both}
\end{equation}

We estimate $\lan |\Bot_x u\theta|\ran_B$ differently than the estimate \eqref{eq: w theta B first} on $\lan|w\theta|\ran_B$. Using the same procedure would require pulling the pointwise maximum of $\Bot$ out of the integral, ultimately giving upper bounds on $\F$ that become infinite as these pointwise maxima become infinite, such as in figure \ref{fig: examples}(c). Instead we remove $\theta$ from the integral using the pointwise bound $|\theta|\le1$ holding at large times, an estimate that follows from the extremum principle $0\le T\le1$ \cite{} and the fact that $0\le\tau\le1$ also. Therefore,
\begin{align}
\lan |\Bot_x u\theta|\ran_B
&\le \lan |\Bot_x u|\ran_B \\
&\le \lan(\Bot_x)^2\ran_B^{1/2}\lan u^2\ran_B^{1/2} \\
&\le \left(\delta\iint_\Omega dxdy\,(\Bot_x)^2\right)^{1/2}
	\left(\tfrac{1}{2}\delta^2\lan u_z^2\ran_B\right)^{1/2} \\
&\le \tfrac{\delta^{3/2}}{2\sqrt2}\left[c'\delta^{3/2} \lan u_z^2\ran_B 
	+ \tfrac{1}{c'\delta^{3/2}}\iint_\Omega dxdy\,(\Bot_x)^2 \right],
\label{eq: u theta B}
\end{align}
where the second line follows from the Cauchy--Schwarz inequality, the third from the estimate \eqref{eq: w pointwise} for $u$, and the last from Young's inequality. Adding \eqref{eq: w theta B} to its counterpart for the top boundary layer gives
\begin{equation}
\lan |\Bot_x u\theta|\ran_B+\lan |\Top_x u\theta|\ran_T \le 
	\tfrac{1}{2\sqrt2}\left(c'\delta^3 \lan u_z^2\ran
	+ \tfrac{1}{c'}\iint_\Omega dxdy\left[(\Bot_x)^2+(\Top_x)^2\right] \right).
\end{equation}
The above estimate and its counterpart for the $y$-direction, along with \eqref{eq: w theta both}, can be applied to \eqref{eq: indef 2} to find
\begin{equation}
\left|\lan\theta\bu\bcdot(2\bnabla\tau-a\zhat)\ran\right| 
\le \left(\tfrac{a+2}{2}\right)\left[ \delta^2\lan \tfrac{c'}{2\sqrt2}(u_z^2+v_y^2)
	+ \tfrac{c}{4}w_z^2\ran + \tfrac{1}{4c}\lan\theta_z^2\ran
	+ \tfrac{1}{\delta}\tfrac{1}{\sqrt2 c'}\hsq A\right],
\label{eq: indef 3}
\end{equation}
where $A=|\Omega|$ is the horizontal area and 
\begin{equation}
\hsq := \tfrac{1}{2A}\iint_\Omega dxdy
	\left(|\bnabla\Bot|^2 + |\bnabla\Top|^2\right)
\label{eq: hsq}
\end{equation}
is the mean squared integral of the boundary gradients. (Note that these gradients have no $z$-components.) To take advantage of the inequality $\lan|u_z|^2+|v_z|^2+2|w_z|^2\ran\le\lan|\bnabla\bu|^2\ran$ that follows from incompressibility and no-slip boundary conditions \citep[cf.][endnote 9]{Constantin1994}, we choose $c'\equiv c/2\sqrt2$ and obtain
\begin{equation}
\left|\lan\theta\bu\bcdot(2\bnabla\tau-a\zhat)\ran\right| 
\le \left(\tfrac{a+2}{2}\right)\left[ \delta^2\tfrac{c}{8}\lan|\bnabla\bu|^2\ran 
	+ \tfrac{1}{4c}\lan|\bnabla\theta|^2\ran 
	+ \tfrac{1}{\delta}\tfrac{2}{c}\hsq A \right],
\label{eq: indef 4}
\end{equation}
where we have also used $\lan\theta_z^2\ran\le\lan|\bnabla\bu|^2\ran$. The above expression is a bound on the sign-indefinite term of $\Q_1$ involving its two sign-definite terms. Applying this estimate to the definition \eqref{eq: Q1} of $\Q_1$ gives
\begin{equation}
\Q_1 \le \left[(1-\gamma)\tfrac{a}{Ra}-c\,\delta^2\left(\tfrac{a+2}{16}\right)\right]
	\lan|\bnabla \bu|^2\ran 
	+ \left[ 1-\tfrac{1}{c}\left(\tfrac{a+2}{8}\right) \right]\lan|\bnabla \theta|^2\ran
	+ \tfrac{1}{\delta}\tfrac{1}{c}(a+2)\hsq A.
\label{eq: Q with coefficients}
\end{equation}
We choose $c$ and $\delta$ to make the above coefficients on $\lan|\bnabla \bu|^2\ran$ and $\lan|\bnabla \theta|^2\ran$ nonpositive, in which case $\Q_1$ is bounded above by the final term, which has no dependence on the unknown fields $\bu$ or $\theta$. Our eventual upper bound on $\F$ is minimized by choosing the largest possible $\delta$. This is done by choosing $c$ as small as possible while the $\lan|\bnabla \theta|^2\ran$ coefficient remains nonpositive,
\begin{equation}
c=\tfrac{a+2}{8},
\end{equation}
and then choosing $\delta$ as large as possible while the $\lan|\bnabla \bu|^2\ran$ coefficient remains nonpositive,
\begin{equation}
\delta = \sqrt{1-\gamma}\left(\tfrac{8\sqrt{2a}}{a+2}\right)Ra^{-1/2}.
\label{eq: delta}
\end{equation}
When $Ra$ is small this formula produces a value of $\delta$ that is too large to properly define $\tau$ because the boundary layers would overlap, so our main result will apply only for sufficiently large $Ra$. As discussed at the end of \S\ref{sec: main result}, this restriction can be lifted by defining $\delta$ differently at small $Ra.$

With the $c$ and $\delta$ chosen above, $\Q_1$ is bounded by the last term of \eqref{eq: Q with coefficients}, which becomes
\begin{equation}
\Q_1 \le \tfrac{1}{\sqrt{1-\gamma}}\left(\tfrac{a+2}{\sqrt{2a}}\right)\hsq ARa^{1/2}.
\label{eq: Q1 final}
\end{equation}
This bound on $\Q_1$ holds for the $\tau$ defined by \eqref{eq: tau} with $\delta$ as in \eqref{eq: delta} and any $a>0$ and $\gamma\in(0,1)$. Thus we have shown that one of the three terms comprising $\F$ in \eqref{eq: F final} is no larger than ${\cal O}(Ra^{1/2})$. The next two subsections show the same for the other two terms, hence $\F$ itself is no larger than ${\cal O}(Ra^{1/2})$. After all three terms have been estimated, particular values of $a$ and $\gamma$ will be chosen to achieve an explicit prefactor that is as small as possible.


\subsection{\label{sec: Q2}Upper bound on $\Q_2$}

In this subsection the quadratic functional $\Q_2[\bu]$ defined by (\ref{eq: Q2}) is bounded above by an expression that is smaller than ${\cal O}(Ra^{1/2})$. We first show that the sign-indefinite term $\lan\tau w\ran$ is small by exploiting the thinness of the boundary layers of $\tau$. The fact that $\tau$ is nonzero outside the boundary layers is not an obstacle because $\lan\tau w\ran=\lan\wt\tau w\ran$, where $\wt\tau:=\tau-\left[\tfrac{1}{2}+\tfrac{a}{2}\left(z-\tfrac{1}{2}\right)\right]$ \emph{does} vanishes outside the boundary layers. The equality $\lan\tau w\ran=\lan\wt\tau w\ran$ holds because $\lan gw\ran=0$ for any function $g(z)$ when the boundaries are no-slip. (This follows from the identity $\lan gw\ran = \lan \bnabla\bcdot(\bu G)\ran$, where $G(z)$ is an antiderivative of $g(z)$, since $\lan \bnabla\bcdot(\bu G)\ran=0$ with impenetrable boundaries.)

Because $\wt\tau$ vanishes outside the boundary layers, $\lan\tau w\ran$ can be expressed in terms of integrals over the boundary layers as
\begin{equation}
\lan\tau w\ran = \lan\wt\tau w\ran_B + \lan\wt\tau w\ran_T.
\label{eq: wt equality}
\end{equation}
Using the pointwise bound \eqref{eq: w pointwise} on $w$ in the bottom boundary layer gives
\begin{align}
\left|\lan\wt\tau w\ran_B\right|
&\le \lim_{t\to\infty}\frac{1}{t}\int_0^tdt'\iint_\Omega dxdy
	\left(\int_{\Bot}^{\Bot+\delta}dz\,w_z^2\right)^{1/2} 
	\left\vert\int_{\Bot}^{\Bot+\delta}dz\,(z-\Bot)^{1/2}\,\wt\tau \right\vert \\
&\le \left(\tfrac{a+2}{15}\right)\delta^{3/2}\lim_{t\to\infty}\frac{1}{t}\int_0^tdt'
	\iint_\Omega dxdy
	\left(\int_{\Bot}^{\Bot+\delta}dz\,w_z^2\right)^{1/2} \\
&\le \left(\tfrac{a+2}{15}\right)\delta^{3/2}A^{1/2}\lan w_z^2\ran_B^{1/2} \\
&\le \tfrac{4\gamma}{Ra}\lan w_z^2\ran_B + \tfrac{Ra}{16\gamma}\left(\tfrac{a+2}{15}\right)^2\delta^{3}A
\label{eq: tau w B}
\end{align}
The second line above is reached after bounding the $z$-integral of $(z-\Bot)^{1/2}\,\wt\tau$, the horizontal area $A$ appears in the third line after using the Cauchy--Schwarz inequality, and in the final line we have used Young's inequality with the intent of making the first coefficient $4\gamma/Ra$. Applying \eqref{eq: tau w B} and its counterpart in the top boundary layer to \eqref{eq: wt equality} gives
\begin{align}
\left|\lan\tau w\ran\right|
&\le \tfrac{4\gamma}{Ra}\lan w_z^2\ran 
	+ \tfrac{Ra}{8\gamma}\left(\tfrac{a+2}{15}\right)^2\delta^{3}A \\
&\le \tfrac{\gamma}{Ra}\lan|\bnabla\bu|^2\ran
	+ \tfrac{(1-\gamma)^{3/2}}{\gamma}\tfrac{128\sqrt2}{225}
	\tfrac{a^{3/2}}{a+2}A Ra^{-1/2},
\end{align}
where the second line has used the definition \eqref{eq: delta} of $\delta$ and the estimate $\lan|w_z|^2\ran\le\tfrac{1}{4}\lan|\bnabla\bu|^2\ran$ that follows from incompressibility and no-slip boundary conditions \citep[cf. (5.16) of][]{Constantin1996}. The functional $\Q_2[\bu]$ defined by \eqref{eq: Q2} is thus bounded by
\begin{equation}
\Q_2 \le \tfrac{1}{\gamma}\tfrac{128\sqrt2}{225}\tfrac{a^{5/2}}{a+2}A Ra^{-1/2}.
\label{eq: Q2 final}
\end{equation}
For convenience we have replaced $(1-\gamma)^{3/2}$ with $1$ in the bound since this does not worsen the the leading prefactor of our eventual bound on $\F$. In \S\ref{sec: gamma} an expression is chosen for $\gamma$ that is ${\cal O}(Ra^{-1/2})$. This makes the above estimate $O(1)$ in $Ra$, so $\Q_2$ makes only a higher-order contribution to our ${\cal O}(Ra^{1/2})$ bound on $\F$.


\subsection{\label{sec: UB on UB}Upper bound on $\lan|\bnabla\tau|^2\ran$}

In this subsection we bound $\lan|\bnabla\tau|^2\ran$, having already bounded the other two terms comprising $\F$ in \eqref{eq: F final}. If a particular geometry is of interest one can evaluate $\lan|\bnabla\tau|^2\ran$ for the pertinent $\Top $ and $\Bot $ and then tune $a$ to minimize the bound. This idea is revisited in \S\ref{sec: sinusoidal}, but first we take the more general approach of bounding $\lan|\bnabla\tau|^2\ran$ above in terms of $\bnabla \Top $ and $\bnabla \Bot $. In this way we establish for \emph{all} admissible geometries that $\lan|\bnabla\tau|^2\ran$ is no larger than ${\cal O}(Ra^{1/2})$.

The horizontal derivative $\tau_x$ vanishes outside the boundary layers and is bounded by (\ref{eq: dx ineq}) in the top boundary layer, and analogously in the bottom one, so its square integral is bounded by 
\begin{equation}
\lan\tau_x^2\ran \le
	\tfrac{1}{\delta}\tfrac{(a+2)^2}{16}\iint_\Omega dxdy
	\left[(\Bot_x)^2+(\Top_x)^2\right].
\label{eq: tau_x}
\end{equation}
The vertical derivative $\tau_z$ is $a/2$ in the interior and is bounded by $(a+2)/4\delta$ in both boundary layers, so
\begin{align}
\lan\tau_z^2\ran &\le \tfrac{1}{\delta^2}\tfrac{(a+2)^2}{16}V_{BL} + \tfrac{a^2}{4}(V-V_{BL}) \\
	&\le \tfrac{1}{\delta}\tfrac{(a+2)^2}{8}A + \tfrac{a^2}{4}V,
\label{eq: tau_z}
\end{align}
where $V$ is the volume of the domain, $A=|\Omega|$ is its cross-sectional area, and $V_{BL}=2A\delta$ is the combined volume of the boundary layers. Using \eqref{eq: tau_z} along with \eqref{eq: tau_x} and its analogue for $\lan\tau_y^2\ran$ gives
\begin{equation}
\lan|\bnabla\tau|^2\ran \le \tfrac{1}{\delta}\tfrac{(a+2)^2}{8}\left(1+\hsq\right)A + \tfrac{a^2}{4}V,
\label{eq: grad tau bound}
\end{equation}
where the mean squared boundary gradient $\hsq$ is as defined by \eqref{eq: hsq}. For the $\delta$ we defined by \eqref{eq: delta} when showing $\Q_1\lesssim {\cal O}(Ra^{1/2})$, the above expression becomes
\begin{equation}
\lan|\bnabla\tau|^2\ran \le \tfrac{1}{\sqrt{1-\gamma}} \tfrac{(a+2)^3}{64\sqrt{2a}}
	\left(1+\hsq\right)A\,Ra^{1/2} 
	+ \tfrac{a^2}{4}V
\label{eq: grad tau bound 2}
\end{equation}
for all $a>0$ and $\gamma\in(0,1)$.


\subsection{\label{sec: main result}Main result}

The three terms comprising $\F$ in expression \eqref{eq: F final} have now been bounded above by \eqref{eq: Q1 final}, \eqref{eq: Q2 final}, and \eqref{eq: grad tau bound 2}. Combining these estimates gives
\begin{equation}
\F \le \tfrac{1}{\sqrt{1-\gamma}}C_1A\,Ra^{1/2}
	+ \tfrac{1}{\gamma}C_2A\,Ra^{-1/2} + \tfrac{a^2}{4}V
\label{eq: F bound 2}
\end{equation}
for all $a>0$ and $\gamma\in(0,1)$, where
\begin{align}
C_1 &:= \tfrac{(a+2)^3}{64\sqrt{2a}} + \tfrac{(a+2)^3+64(a+2)}{64\sqrt{2a}} \hsq, &
C_2 &:= \tfrac{128\sqrt2}{225}\tfrac{a^{5/2}}{a+2}.
\label{eq: C1}
\end{align}
The optimal $\gamma$ that minimizes \eqref{eq: F bound 2} depends on $Ra$, and at large $Ra$ this optimal value asymptotes to
\begin{equation}
\gamma = \sqrt{\tfrac{2C_2}{C_1}}Ra^{-1/2}.
\label{eq: gamma}
\end{equation}
With this choice of $\gamma$, the bound \eqref{eq: F bound 2} on $\F$ becomes
\begin{equation}
\F \le \tfrac{1}{\sqrt{1-\sqrt{2C_2/C_1}Ra^{-1/2}}}C_1A\,Ra^{1/2}
	+ \left[ \sqrt{\tfrac{C_1C_2}{2}}A+\tfrac{a^2}{4}V \right]
\label{eq: F bound 3}
\end{equation}
for all $a>0$.

Expression \eqref{eq: F bound 3} is our main result. For \emph{any} domain where the functions defining the top and bottom boundaries have square-integrable gradients, we have proven that the time-averaged heat flux through the layer, $\F$, can grow no faster than $Ra^{1/2}$ at large $Ra$. The particulars of the geometry enter through the total fluid volume $V$, the horizontal area $A$, and the mean squared boundary gradients $\hsq$ defined by \eqref{eq: hsq}.

The bound \eqref{eq: F bound 3} is really a family of results; each choice of $a>0$ gives an explicit bound on $\F$ that varies with $Ra$ and $\hsq$. If the aim is to optimize the bound at large $Ra$, then $a$ should be chosen to minimize $C_1$ because the leading behaviour of \eqref{eq: F bound 3} is
\begin{equation}
\F \lesssim A\left[C_1Ra^{1/2}
	+ {\cal O}(1)\right].
\label{eq: F bound leading}
\end{equation}
The value of $a$ that minimizes $C_1$ depends on the boundary geometry since the definition \eqref{eq: C1} of $C_1$ involves $\hsq$. If a particular geometry is of interest, one can calculate $\hsq$ and then choose $a$. To give a concrete bound here we simply fix $a=2/5$, so
\begin{align}
C_1 &= \tfrac{27}{50\sqrt5} + \tfrac{327}{50\sqrt5}\hsq \approx 0.242 + 2.925\hsq, \\
C_2 &= \tfrac{256}{3375\sqrt5} \approx 0.0339.
\label{eq: C small h}
\end{align}
This choice of $a$ minimizes $C_1$ when the boundaries are planar ($\hsq=0$), and even in the worst case where $\hsq\gg1$ it gives a value for $C_1$ that is no more than 23\% larger than optimal. Applying the above $C_1$ and $C_2$ to \eqref{eq: F bound 3} gives an explicit bound on $\F$ whose leading behaviour is
\begin{equation}
\F \lesssim A\left[\left(\tfrac{27}{50\sqrt5} + \tfrac{327}{50\sqrt5}\hsq\right)Ra^{1/2} + {\cal O}(1)\right].
\label{eq: 2/5 explicit}
\end{equation}
With planar boundaries the leading prefactor reduces to 0.242, which is unsurprisingly larger than the best known prefactor of 0.027 proven specifically for planar boundaries by \citet{Plasting2003}, although not very much larger than the prefactor of 0.167 found by \citet{Constantin1996}.

The bound \eqref{eq: F bound 3} holds only for sufficiently large $Ra$ because our analysis has required that $\gamma<1$ and that the boundary layers of $\tau$ fit in the domain, meaning $2\delta\le \Delta_h:=\min_{x,y}(\Top -\Bot)$. For the explicit bound \eqref{eq: 2/5 explicit} where we fixed $a=2/5$, these conditions hold if
\begin{equation}
Ra > \max\left\{\tfrac{1024}{3645},\tfrac{320}{9}\tfrac{1}{\Delta_h^2}\right\}.
\label{eq: min Ra}
\end{equation}
With planar boundaries this becomes $Ra>320/9\approx35.6$, which is an irrelevant restriction since convection cannot persist if $Ra<1707.7$. With non-planar boundaries, the occurrence of convection for all $Ra>0$ means that bounds on $\F$ may be wanted at small $Ra$. The restriction \eqref{eq: min Ra} on $Ra$ arose only because we chose $\delta$ and $\gamma$ to be optimal at large $Ra$, and it is straightforward to modify our analysis to prove a bound holding for all $Ra>0$. When $Ra$ is too small for \eqref{eq: gamma} to obey $\gamma<1$ one can instead fix a constant value of $\gamma$, and when $Ra$ is too small for boundary layers defined by \eqref{eq: delta} to obey $2\delta\le\Delta_h$, one can instead take $2\delta=\Delta_h$.


\subsection{\label{sec: sinusoidal} Sinusoidal boundaries}

Consider the example of sinusoidal boundaries, a domain recently simulated in two dimensions by \citet{Toppaladoddi2015}. Suppose the top and bottom boundaries are uniform in $y$ and sinusoidal in $x$ with amplitude $\beta$ and wavelength $\lambda$. Let the span in the $x$-direction be a multiple of $\lambda$, so an integer number of wavelengths fit in the domain. The flow may be either horizontally periodic or bounded by vertical side walls. The boundary gradients $\Bot_x$ and $\Top_x$ equal $\tfrac{2\pi \beta}{\lambda}\sin\left( \tfrac{2\pi x}{\lambda} \right)$ or translations thereof, so the mean squared boundary gradient is $\hsq = 2\pi^2\beta^2/\lambda^2$. Expression \eqref{eq: 2/5 explicit} for the leading behaviour of the bound then becomes
\begin{equation}
\F \lesssim A\left[ \left(\tfrac{27}{50\sqrt5} 
	+ \tfrac{327\pi^2}{25\sqrt5}\tfrac{\beta^2}{\lambda^2}\right)Ra^{1/2} 
	+ {\cal O}(1)\right].
\label{eq: F sinusoidal}
\end{equation}
It is similarly easy to evaluate \eqref{eq: 2/5 explicit} for other geometries, but we can improve upon this general result by specializing the analysis at an earlier stage.

To illustrate how to improve upon our main result for a particular geometry of interest, let us consider sinusoidal boundaries whose wavelength is twice their amplitude, meaning $\lambda=2\beta$ and so $\hsq = \pi^2/2$. The bound \eqref{eq: F sinusoidal} then has a leading prefactor of about 14.67. This bound exceeds the heat flux in the computations of \citet{Toppaladoddi2015} by about two orders of magnitude, suggesting room for improvement. The prefactor of 14.67 is simply the coefficient $C_1$ defined by \eqref{eq: C1}, after we have chosen $a=2/5$ and used the value of $\hsq$, but the choice $a=2/5$ is sub-optimal when $\hsq$ is not small. A simple way to lower the prefactor is to choose $a\approx1.2556$, which minimizes $C_1$ when $\hsq = \pi^2/2$. This lowers the leading prefactor to about 12.15, but we can do still better by revisiting the analysis that led to the expression for $C_1$.

The leading prefactor of our main result \eqref{eq: F bound 3} is the sum of the leading coefficients of the bound \eqref{eq: Q1 final} on $\Q_1$ and the bound \eqref{eq: grad tau bound 2} on $\lan|\bnabla\tau|^2\ran$. One way to tighten the result is to evaluate $\lan|\bnabla\tau|^2\ran$ exactly instead of bounding it above. Doing so in the case of sinusoidal boundaries with $\lambda=2\beta$ gives a bound on $\F$ with a lengthy leading prefactor that depends on $\beta$ and $a$. For each value of $\beta$ one can choose the value of $a$ that minimizes the prefactor. Doing so recovers the prefactor of 12.15 when $\beta\ll1$ but gives a better result as $\beta$ increases. When $\beta=0.2$, for instance, the optimal choice $a\approx1.431$ gives a bound on $\F$ with a leading prefactor of about 11.56.

A final way to tighten our main result in some cases is to revisit the bound on $\Q_1$, where the term $\lan|2\tau_x u\theta|\ran$ was estimated by \eqref{eq: u theta B}. In deriving this estimate, the pointwise bound $|\theta|\le1$ was used to remove $\theta$ from the integral. One could instead remove the pointwise maxima of $\Bot_x$ and $\Top_x$ from the integral and then estimate $\lan|u\theta|\ran$ in the boundary layer in the same way that $\lan|w\theta|\ran$ was estimated by \eqref{eq: w theta B}. Carrying through this analysis gives an ${\cal O}(Ra^{1/2})$ upper bound on $\F$ with a leading prefactor that involves not only the mean squared values of $\left|\bnabla\Bot\right|$ and $\left|\bnabla\Top\right|$ but also their pointwise maxima. Such a result is generally weaker than our main result \eqref{eq: F bound 3} since it requires boundary gradients to be uniformly bounded, but it gives smaller prefactors when the pointwise maxima are fairly small. In example of sinusoidal boundaries with $\lambda=2\beta$, this alternate analysis gives leading prefactors that are about half as large as in our present analysis.


\section{Conclusions \label{sec: con}}

Not unexpectedly, it is more work to bound quantities of physical interest in complicated geometries than in the idealized spatial domains of canonical models. The effort is nonetheless worthwhile since all real geometries are imperfect, and even small-scale wall roughness can significantly alter mean quantities. Here we have shown how to adapt the background method to establish bounds in complicated geometries. Almost all previous applications of this method have been limited to simple geometries, with the notable exception of \citet{Wang1997}. The configuration we have studied is Rayleigh--B\'enard convection in a fluid layer whose top and bottom boundary heights are functions of horizontal position. Our main finding is that heat transport across the fluid layer, normalized by the temperature difference between the boundaries, is ${\cal O}(Ra^{1/2})$ at large $Ra$. This is the same scaling of previously proven bounds for finite-Prandtl-number convection between no-slip \emph{planar} boundaries. Our bound is consistent with past numerical and laboratory studies of convection between rough boundaries, where growth of the Nusselt number has been either proportional to $Ra^{1/2}$ \citep{Roche2001} or slower.

A remaining challenge is to remove some of the technical assumptions needed here. Our proof requires the boundaries heights to be functions of horizontal position with square-integrable gradients. This is because we defined the boundary layers of the thermal background field to have a fixed vertical thickness. More general boundary geometries might be tackled by instead defining boundary layers according to their thickness normal to the boundaries. Furthermore, our requirement that side boundaries be planar and vertical might be relaxed with additional effort.

A more ambitious goal for future analysis is to extend the background method to complicated geometries when a background decomposition must be applied to the velocity field, as opposed to a scalar field like the temperature. Such velocity decompositions have been used to prove upper bounds on viscous dissipation in canonical models of shear flows, where simple geometry allows for plane parallel background flows that automatically satisfy incompressibility. In complicated geometries, such as for channel flow between rough walls, constructing suitable incompressible background flows presents difficulties that are the subject of current research.

\newpage

\medskip
\noindent
{\it Acknowledgements.}
Much of this research was performed at the Woods Hole Oceanographic Institution's Geophysical Fluid Dynamics Program, substantially supported by US National Science Foundation (NSF) Awards OCE-0824636 and OCE-1332750 and the Office of Naval Research, and at the NSF's Institute for Pure \& Applied Mathematics during its 2014 {\it Mathematics of Turbulence} program. This work was also supported in part by NSF Awards PHY-1205219 and DMS-1515161, a {\it Simons Fellowship in Theoretical Physics} and a {\it John Simon Guggenheim Memorial Foundation Fellowship} (CRD). We are furthermore grateful for helpful suggestions from the anonymous referees.

\bibliographystyle{jfm}
\bibliography{bumpyRBv10.bbl}

\end{document}